\documentclass[final,5p,times,twocolumn,preprint]{elsarticle}
\usepackage{slashed}
\usepackage{eucal} 
\usepackage[hypertexnames=false, colorlinks=true, citecolor=blue, urlcolor=blue, linkcolor=black]{hyperref}
\usepackage{amsmath,amssymb,amsfonts,latexsym}
\newcommand{\fourint}[1]{\int\!\frac{d^4 #1}{(2\pi)^4}}
\allowdisplaybreaks
\begin{document}
\begin{frontmatter}
\title{Instanton effects in twist-3 generalized parton distributions}
\author[j]{June-Young Kim}
\ead{jykim@jlab.org}
\author[j]{Christian~Weiss}
\ead{weiss@jlab.org}
\address[j]{Theory Center, Jefferson Lab, Newport News, VA 23606, USA}
\begin{abstract}
The instanton vacuum picture is used to study hadronic matrix elements of the twist-3
(dimension-4, spin-1) QCD operators measuring the quark spin density and spin-orbit correlations.
The QCD operators are converted to effective operators in the low-energy effective theory emerging
after chiral symmetry breaking, in a systematic approach based on the diluteness of the instanton
medium and the $1/N_c$ expansion. The instanton fields induce spin-flavor-dependent ``potential''
terms in the effective operators, complementing the ``kinetic'' terms from the quark field momenta.
As a result, the effective operators obey the same equation-of-motion relations as the original QCD operators.
The spin-orbit correlations are qualitatively different from naive quark model expectations.
\end{abstract}
\end{frontmatter}
\section{Introduction}
The spontaneous breaking of chiral symmetry is fundamental to the emergence of hadrons from QCD.
It generates most of the light hadron masses through the action of chromodynamic fields in the vacuum.
It produces the pion as a quasi-massless excitation mediating long-range interactions between hadrons.
The effective dynamics at large distances $\sim 1/M_\pi$ is expressed by the chiral Lagrangian,
describing the interaction of pions and their coupling to the massive hadrons.

Hadron structure is expressed in the matrix elements of composite QCD operators between hadronic states.
At the level of the effective dynamics, these QCD operators can be represented by operators in the
effective degrees of freedom. One can distinguish two types of QCD operators: (i) Conserved currents related
to global symmetries, e.g.\ the vector and axial vector currents and the energy-momentum tensor.
For such QCD operators the effective operators can be obtained from the response of the chiral Lagrangian
to the symmetry transformations (Noether theorem). No dynamical information is needed beyond that
contained in the chiral Lagrangian. (ii) QCD operators not related to symmetries. This covers the
vast majority and includes the quark/gluon operators characterizing partonic structure, e.g. the local
twist-2 spin-$n$ and twist-3 spin-$n$ operators determining the moments of the generalized parton
distribution (GPDs). For such QCD operators the derivation of the effective operators requires 
dynamical information beyond what is in the chiral Lagrangian.

Numerous observations suggest that chiral symmetry breaking is caused by topological fluctuations of
the gauge fields in the QCD vacuum (instantons); see Refs.~\cite{Diakonov:2002fq,Schafer:1996wv}
for a review. Lattice simulations of Euclidean (imaginary-time) QCD show that these fluctuations
are localized with average size $\bar\rho \sim$ 0.3 fm and occur at average distances $\bar R \sim$ 1 fm,
and that they involve strong semiclassical fields. These gauge fields induce localized zero modes
of the fermion fields with definite chirality, which delocalize to break chiral symmetry.
The instanton vacuum is abstracted from these findings and describes the QCD gauge fields
on the scale $\bar\rho^{-1}$ as a superposition of instantons, coupled to the fermions
by the zero modes \cite{Diakonov:2002fq,Schafer:1996wv}. The picture is endowed with a
small parameter in the form of the packing fraction of instantons in the vacuum,
$\kappa \equiv \pi^2 \bar\rho^4 / \bar R^4 \approx 0.1$ (diluteness).
Chiral symmetry breaking and hadronic correlation functions can be studied in a systematic expansion.
It leads to a successful phenomenology of hadron structure, validated by many observations.
Using techniques based on the $1/N_c$ expansion (saddle point approximation, bosonization)
one can construct the effective dynamics at the scale $\bar R$, described by quarks with a
dynamical mass $M \sim$ 0.3--0.4 GeV, coupled to a chiral
pion field \cite{Diakonov:1985eg,Diakonov:1995qy,Kacir:1996qn}.
The chiral Lagrangian is obtained by integrating out the quarks and performing a gradient
expansion in the pion field.

In the instanton vacuum one can derive the effective operators representing QCD operators in
the effective theory after chiral symmetry breaking \cite{Diakonov:1995qy}.
The picture provides an explicit model of the
nonperturbative gauge fields appearing in quark-gluon and pure gluon QCD operators.
The effective operators can be derived systematically in an expansion in the packing fraction.
The method has been used to analyze and predict the nucleon matrix elements of several
higher-twist quark-gluon operators \cite{Balla:1997hf,Dressler:1999zi}
and higher-dimensional gluon operators \cite{Weiss:2021kpt}.
It preserves operator relations following from QCD equations of motion \cite{Balla:1997hf},
as well as low-energy theorems from the scale and $U(1)_A$ anomalies of QCD \cite{Diakonov:1995qy}.

In this work we use the instanton vacuum to study the hadronic matrix elements of the twist-3
QCD operators measuring the quark spin density and spin-orbit correlations in hadrons.
These operators appear in the theory of twist-3 generalized parton distributions (GPDs) and the
phenomenology of nucleon spin structure. We derive the effective operators and find that
parametrically large ``potential'' terms arise from the instanton gauge field in the covariant
derivatives, complementing the ``kinetic'' terms from the quark field momenta. As a consequence,
the effective operators obey the same equation-of-motion relations as the original QCD operators,
a very gratifying result. We discuss the quark spin-orbit correlations
described by the instanton-based effective operators and show that they are qualitatively different
from those obtained with quark model-based operators including only kinetic terms.
\section{Twist-3 QCD operators}
We consider the twist-3 (dimension-4, spin-1) operators
\begin{align}
O^{\alpha\beta}(x) &\equiv \frac{1}{2}
\bar\psi (x) \gamma^{[\alpha} i \overleftrightarrow{\nabla}^{\beta]} \, \tau \, \psi (x),
\label{operator_flavor_natural}
\\
O_{5}^{\alpha\beta}(x) &\equiv \frac{1}{2}
\bar\psi (x) \gamma^{[\alpha} \gamma_5 \, i \overleftrightarrow{\nabla}^{\beta]}
\, \tau \, \psi (x),
\label{operator_flavor_unnatural}
\end{align}
with $[\alpha\beta] \equiv \alpha\beta - \beta\alpha$. Here
\begin{align}
\overleftrightarrow{\nabla}^\beta
&\equiv \frac{1}{2}
(\overrightarrow{\partial}^\beta - \overleftarrow{\partial}^\beta )
- i \frac{\lambda^c}{2} A^{c\beta}(x)
\end{align}
is the QCD covariant derivative with gauge potential $A^{c\beta}(x)$. $\psi(x)$ is the quark field
with $N_f$ light flavors; $\tau$ denotes a generic flavor matrix;
both singlet ($\tau = 1$) and non-singlets ($\tau = \tau^a$) will be discussed in following.
The natural-parity operator Eq.~(\ref{operator_flavor_natural}) represents the antisymmetric part of
the QCD energy-momentum tensor and measures the quark spin density in hadrons \cite{Lorce:2017wkb}.
The unnatural-parity operator Eq.~(\ref{operator_flavor_unnatural}) describes
quark spin-orbit correlations \cite{Lorce:2014mxa}.
Using the QCD equations of motion, the operators can be expressed in terms of the twist-2 axial vector
and vector current operators,
\begin{align}
\frac{1}{2}
\bar\psi (x) \gamma^{[\alpha} i \overleftrightarrow{\nabla}^{\beta]} \, \tau \, \psi (x)
&= -\frac{1}{4}
\epsilon^{\alpha\beta\gamma\delta} \partial_\gamma \left[\psi (x) \gamma_\delta \gamma_5 \tau \psi (x) \right],
\label{eom_natural}
\\
\frac{1}{2}
\bar\psi (x) \gamma^{[\alpha} \gamma_5 i \overleftrightarrow{\nabla}^{\beta]} \, \tau \, \psi (x)
&= -\frac{1}{4}
\epsilon^{\alpha\beta\gamma\delta} \partial_\gamma \left[\psi (x) \gamma_\delta \tau \psi (x) \right]
+ ...,
\label{eom_unnatural}
\end{align}
where $\partial_\gamma [...]$ denotes the total derivative. In Eq.~(\ref{eom_unnatural})
we have omitted operators proportional to the current quark masses. In our convention $\epsilon^{0123} = 1$,
$\gamma_5 \equiv -i\gamma^0\gamma^1\gamma^2\gamma^3$ \cite{LLIV}. Equations~(\ref{eom_natural})
and (\ref{eom_unnatural}) imply that the hadronic matrix elements of the operators
$\langle p_2 | ... | p_1 \rangle$
are proportional to the 4-momentum transfer $p_2 - p_1$ and vanish in the forward limit.

The scale dependence of the twist-3 operators can be inferred from Eqs.~(\ref{eom_natural}) and (\ref{eom_unnatural}).
The vector currents and the flavor-nonsinglet axial currents are conserved and scale-independent;
the flavor-singlet axial current is subject to the $U(1)_A$ anomaly.
\section{Effective dynamics from instantons}
The foundations of the instanton vacuum are described in detail in Refs.~\cite{Diakonov:2002fq,Schafer:1996wv}; 
here we only introduce the elements used in the present calculation, following
the formulation of Ref.~\cite{Diakonov:1995qy}. From the Minkowskian coordinates $(x^0, x^k)$
$(k = 1, 2, 3)$, Euclidean coordinates are defined as $x_\alpha = (x_k, x_4) \equiv (x^k, ix^0)$,
with norm $x^2 = x_\alpha x_\alpha \equiv x_k^2 + x_4^2$. Euclidean derivatives
are defined as $\partial_\alpha = (\partial_k, \partial_4) \equiv (-\partial^k, -i\partial^0)$;
the gauge potential is $A_\alpha = (A_k, A_4) \equiv (-A^k, -iA^0)$,
and the covariant derivative is $\nabla_\alpha = \partial_\alpha - i A_\alpha$.
Hermitean gamma matrices are introduced as $\gamma_\alpha = (\gamma_k, \gamma_4)
\equiv (-i \gamma^k, \gamma^0)$. The Euclidean $\gamma_5$ matrix is
$\gamma_5 \equiv \gamma_1 \gamma_2 \gamma_3 \gamma_4$ and numerically the same as the Minkowskian $\gamma_5$.
The conjugate fermion field is taken as $\psi^\dagger \equiv i \bar\psi$.

In the instanton vacuum the QCD gauge potential is represented by a sum of instanton and
antiinstanton ($I$ and $\bar I$) fields in singular gauge, and the functional integral is
performed over the collective coordinates (position, color orientation, size).
Large instanton sizes are suppressed by instanton interactions, and an effective size distribution
centered around $\bar\rho \sim$ 0.3 fm is obtained. Fluctuations of the size are
suppressed by $1/N_c$, and one can take $\rho = \bar\rho$ in leading order
\cite{Diakonov:1983hh,Diakonov:1995qy}.

The quark field modes with Euclidean momenta $p \lesssim \bar\rho^{-1}$ experience chiral symmetry
breaking and can be described by an effective field theory. Each $I (\bar I)$ induces a
multifermion vertex of the form of a flavor determinant (see Fig.~\ref{fig:effop}a)
\begin{align}
\propto \; \textrm{det}_{f'f} \, {\psi^\dagger}_{\!\!\! f'} (z) \, F(\overleftarrow{\partial})
\frac{1 \pm \gamma_5}{2}
F(\overrightarrow{\partial}) \,
\psi_f (z),
\end{align}
where $z$ is the $I(\bar I)$ position; the color orientation has been integrated over.
We use the shorthand notation
\begin{align}
F(\overrightarrow{\partial}) \, \psi(z) 
&\equiv \fourint{p} e^{i p \cdot z} F(p) \psi (p),
\\
\psi^\dagger (z) \, F(\overleftarrow{\partial})
&\equiv \fourint{p'} e^{-i p' \cdot z} \psi^\dagger (p') F(p').
\end{align}
$F(p)$ is the wave function of the $I (\bar I)$ fermionic zero mode with range $p \lesssim \bar\rho^{-1}$,
normalized such that $F(0) = 1$ \cite{Diakonov:1995qy}.
The ground state of the interacting fermion system is constructed in the $1/N_c$ expansion
(saddle point approximation) \cite{Diakonov:1985eg,Diakonov:1995qy}. A non-trivial saddle point appears,
characterized by a dynamical quark mass $M$ of parametric order $M^2 \sim \kappa \bar\rho^{-2}$ and
numerical value $M \sim$ 0.3--0.4 GeV. The functional integral is
bosonized by introducing a chiral field (see Fig.~\ref{fig:effop}b)
\begin{align}
U(x) \equiv \exp [i \pi^a (x) \tau^a/F_\pi],
\end{align}
where $F_\pi$ defines the normalization of the pion field. This converts the multifermion interaction
to a Yukawa-type interaction of the quarks with the chiral field. At the saddle point, the effective
action takes the simple form
\begin{align}
S_{\rm eff}(x) &= \int d^4 \! x \, \psi^\dagger(x) \left[ -i \slashed{\partial}
- i M F(\overleftarrow{\partial}) U^{\gamma_5}(x) F(\overrightarrow{\partial})
\right] \psi (x) ,
\label{S_eff}
\\
U^{\gamma_5}(x) &\equiv \exp \left[ \frac{i \pi^a (x) \tau^a \gamma_5}{F_\pi}\right]
= \frac{1 + \gamma_5}{2} U(x) + \frac{1 - \gamma_5}{2} U^\dagger(x). \!\!\!
\label{U_gamma_5}
\end{align}
Hadronic correlation functions can be computed systematically in the $1/N_c$ expansion
\cite{Diakonov:1985eg,Diakonov:1995qy,Kacir:1996qn}. Baryon correlation functions are
characterized by a non-trivial classical chiral field (soliton),
giving rise to a rich structure \cite{Diakonov:1987ty}. The applications of this model have been
discussed extensively in the literature; the present study focuses on operators in the
theory defined by Eq.(\ref{S_eff}) and does not require details of the correlation functions.

In the same scheme of approximations one can integrate out the quark fields and derive the effective
action of the chiral field,
\begin{align}
& W_{\rm eff}[U]
= \log \textrm{Det} [-i \slashed{\partial} - i M F U^{\gamma_5} F]
\nonumber
\\
&= \int d^4 x \; \left\{ \frac{F_\pi^2}{4}
\textrm{tr} \, [\partial_\alpha U^\dagger \partial_\alpha U] + \mathcal{O}(\partial U^4)
\right\} .
\label{chiral_action}
\end{align}
The dependence on the chiral field can be made explicit by gradient expansion in $\partial U$.
The integrand in Eq.~(\ref{chiral_action}) is the chiral Lagrangian. The low-energy constants
such as $F_\pi^2$ are given by quark loop integrals in the effective theory of Eq.~(\ref{S_eff})
and calculable in terms of the dynamical
scales $M$ and $\bar\rho$. In this way chiral symmetry breaking by instantons quantitatively
predicts the effective dynamics at the hadronic scale.
\section{Effective operators from instantons}
%
% FIGURE
%
\begin{figure}[t]
\centering
\includegraphics[width=.75\columnwidth]{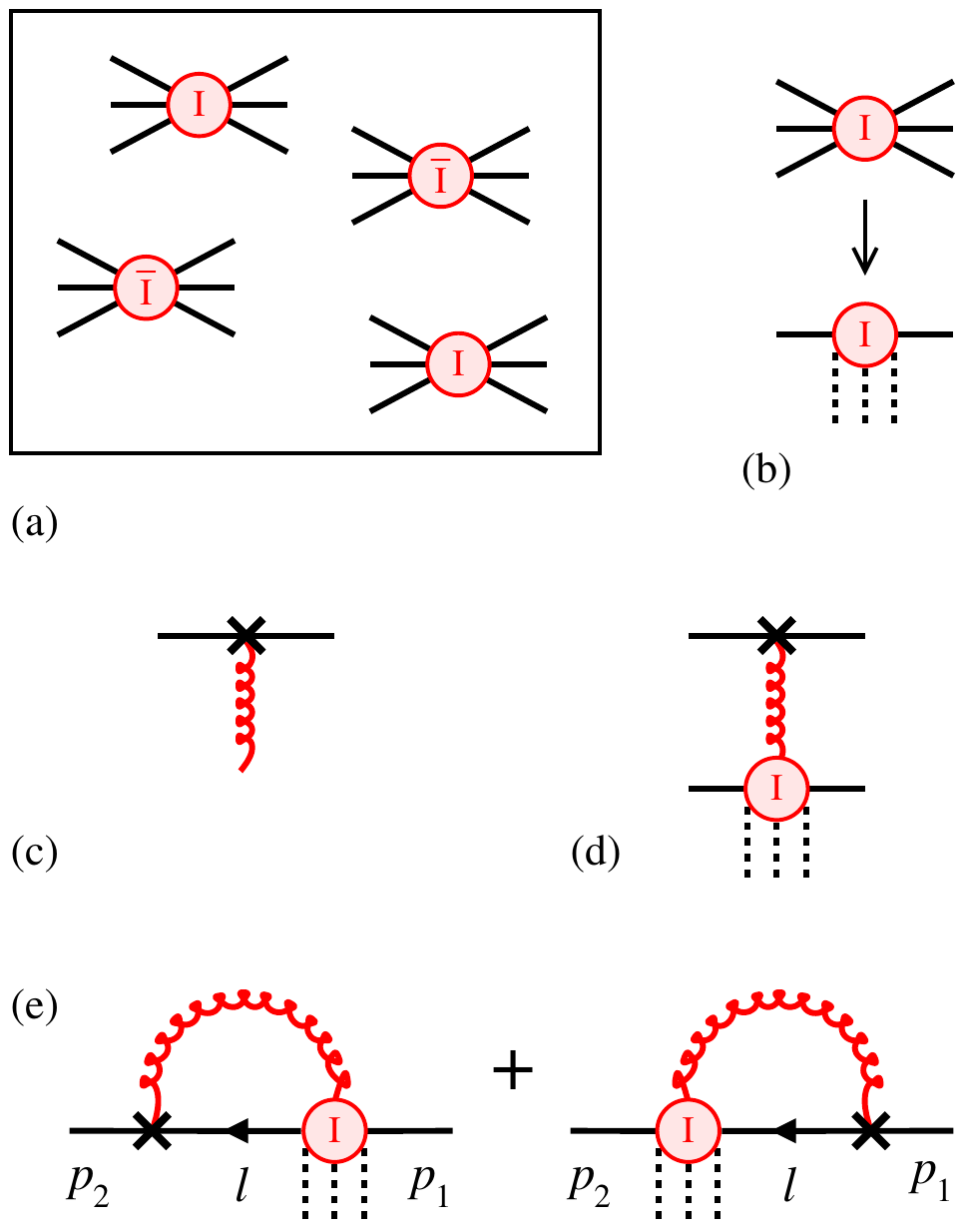} 
\caption{Instanton vacuum and effective operator formalism.
(a)~Multifermion interactions in instanton vacuum. (b)~Bosonization of multifermion interaction.
Dashed lines denote the chiral background field $U(x)$.
(c)~QCD quark-gluon operator. (d)~Effective operator induced by instanton.
(e)~Correlation function of effective operator with quark fields of momenta $p_{1, 2} \sim M$.
The parametrically large contribution arises from the contractions shown here.}
\label{fig:effop}
\end{figure}
We now derive the effective operators for the twist-3 QCD operators Eqs.~(\ref{operator_flavor_natural})
and (\ref{operator_flavor_unnatural}) in instanton vacuum, in the formulation of
Refs.~\cite{Diakonov:1995qy,Balla:1997hf}. To cover all cases of interest, we consider a
general Euclidean operator
\begin{align}
(O_\Gamma)_{\alpha\beta}(x)
&\equiv -i \psi^\dagger (x) \, \Gamma_{\alpha} \overleftrightarrow{\nabla}_\beta \tau \, \psi (x),
\label{operator_euclidean}
\end{align}
with $\Gamma_\alpha \equiv \gamma_\alpha$ or $\gamma_\alpha \gamma_5$ and unsymmetrized tensor
indices $\alpha, \beta$; (anti-) symmetrization will be performed in the expressions below.
The Euclidean operator Eq.~(\ref{operator_euclidean}) is defined such that its spatial components coincide
with those of the Minkowskian operators Eqs.~(\ref{operator_flavor_natural}) and (\ref{operator_flavor_unnatural}),
$\frac{1}{2} (O_\Gamma)_{[ij]} = O^{ij}, O_5^{ij} \, (i,j = 1,2,3)$. The gauge potential-dependent part
of Eq.~(\ref{operator_euclidean}) is (see Fig.~\ref{fig:effop}c)
\begin{align}
(O_\Gamma)_{\alpha\beta}(x)[\textrm{$A$-dep}]
&= - \psi^\dagger (x) \Gamma_{\alpha} \frac{\lambda^c}{2} \tau \, \psi (x) \, A^c_{\beta}(x).
\label{operator_euclidean_potential}
\end{align}
In leading order of the packing fraction, the gauge potential is evaluated in the field of a single
$I(\bar I)$. For an $I(\bar I)$ centered at the origin and in standard color orientation, it is given by
\begin{align}
A^c_\beta (x)_{I (\bar I)} &= (\eta^\mp )^c_{\beta\gamma} \mathcal{A}_{\gamma}(x),
\hspace{2em}
\mathcal{A}_{\gamma}(x) \equiv \frac{2 \bar\rho^2 x_\gamma}{(x^2 + \bar\rho^2 ) x^2},
\label{A_inst}
\end{align}
where $(\eta^\mp)^c_{\alpha\beta} \equiv \bar\eta^c_{\alpha\beta}, \eta^c_{\alpha\beta}$ are
the 't Hooft symbols. The effective operator is constructed
by substituting the $I(\bar I)$ gauge potential in Eq.~(\ref{operator_euclidean_potential}),
multiplying with the zero mode projector, integrating over color orientations, combining the
$I$ and $\bar I$ contributions, and bosonizing the vertex in leading order
of $1/N_c$ \cite{Diakonov:1995qy,Balla:1997hf}. We obtain (see Fig.~\ref{fig:effop}d)
\begin{align}
(O_\Gamma)_{\alpha\beta}(x)
&= - \psi^\dagger (x) \Gamma_{\alpha} \frac{\lambda^c}{2} \tau \psi (x) \;
\frac{i M}{N_c} \int d^4 z \, \mathcal{A}_{\gamma} (x - z) \,
\nonumber \\
& \times \psi^\dagger (z) F(\overleftarrow{\partial}) \frac{\lambda^c}{2}
\sigma_{\beta\gamma} U^{\gamma_5} (z) F(\overrightarrow{\partial}) \psi (z),
\label{effective_operator}
\end{align}
where $\sigma_{\beta\gamma} \equiv (i/2) [\gamma_\beta , \gamma_\gamma ]$. One sees that the
gauge potential has been replaced by a color current of the quark fields coupling to the instanton
zero mode, with a spin-flavor structure dictated by the symmetries of the instanton field.
The effective operator Eq.~(\ref{effective_operator}) represents the QCD
operator Eq.~(\ref{operator_euclidean_potential})
in the bosonized effective theory of Eq.(\ref{S_eff}) within the overall scheme of approximations
(packing fraction, $1/N_c$ expansion).\footnote{Eq.~(\ref{effective_operator})
represents the effective operator for QCD
operators that do not have a vacuum expectation value. It can be used for the spin-2 (symmetric traceless tensor)
and spin-1 (antisymmetric tensor) projections of $(O_\Gamma)_{\alpha\beta}$, but not for the spin-0 projection (trace).
For QCD operators with a vacuum expectation value, there is an additional disconnected contribution
to the effective operator (so-called $\mathcal{R}$ factor) \cite{Balla:1997hf,Diakonov:1995qy}.
\label{footnote:vev}}

The effective operator Eq.~(\ref{effective_operator}) can be inserted in correlation functions
with hadronic currents to extract the hadronic matrix elements. In these correlation functions
the momenta of the external quark fields (coupling to the hadronic currents) are of order
$p \sim M \ll \bar\rho^{-1}$, because the hadronic size is of order $M^{-1}$.
Parametrically large contributions can arise only from loop diagrams in which the fields in the
multifermion operator Eq.~(\ref{effective_operator}) are contracted among themselves
and the internal momenta can extend up to $l \sim \bar\rho^{-1}$ (see Fig.~\ref{fig:effop}e).
This permits further simplification and allows us to reduce the effective operator to a
two-fermion operator of quarks in the chiral background field, similar to the interaction in
Eq.~(\ref{S_eff}).

The spatial variation of the chiral background field $U(z)$ is on the scale $M^{-1} \ll \bar\rho$.
The loop integral with momenta $l \sim \bar\rho^{-1}$ is local on that scale.
In computing the contractions of Eq.~(\ref{effective_operator})
we can therefore neglect the variation of the background field
and set $U =$ const., so that it becomes like a coupling constant.
The contractions can then be computed in momentum representation,
with the momentum assignments of Fig.~\ref{fig:effop}e.
The external momenta are $p_{1, 2} \sim M \ll \bar\rho^{-1}$.
Inside the loop, for the parametrically leading contribution we can neglect the mass term
in the quark propagator with momentum $l$ and replace it by the free propagator $l\cdot \gamma /l^2$.
The momentum representation of the instanton field Eq.(\ref{A_inst}) is
\begin{align}
\int d^4 x\, e^{-ik \cdot x} \; \mathcal{A}_{\gamma} (x)
&= -i \bar\rho^4 \, k_\gamma \, A (k),
\label{potential_fourier_explicit}
\end{align}
where $A(k)$ is a dimensionless scalar function of $k\bar\rho$. The contraction of the
operator Eq.~(\ref{effective_operator}) is obtained as
\begin{align}
\Gamma_{\alpha\beta} &\equiv -\frac{M \bar\rho^4}{2} \fourint{l} \frac{F(l)}{l^2}
\nonumber \\
& \times \left[ A(-l + p_1) (-l + p_1)_\gamma l_\delta \, \Gamma_\alpha \gamma_\delta
\sigma_{\beta\gamma} \, \tau U^{\gamma_5}
\right.
\nonumber
\\
& \left. + \;\; A(-p_2 + l) (-p_2 + l)_\gamma l_\delta \, U^{\gamma_5} \tau \, \sigma_{\beta\gamma} \gamma_\delta
\Gamma_\alpha \right].
\label{contraction}
\end{align}
The zero mode wave functions evaluated at the external momenta have been set to unity, $F(p_{1, 2}) = 1$,
because $p_{1, 2} \ll \bar\rho^{-1}$. A parametrically large contribution arises from the
integral in which the loop momenta in the numerator are projected as
\begin{align}
l_\gamma l_\delta &\rightarrow \delta_{\gamma\delta} \, l^2 / 4.
\end{align}
The factor $l^2$ cancels the denominator of the quark propagator and produces an integral that
would be quadratically divergent if not for the internal zero mode wave function $F(l)$.
Making the projection, simplifying the products of gamma matrices, and dropping
terms $\propto p_{1, 2} \sim M$, Eq.~(\ref{contraction}) becomes
\begin{align}
\!\Gamma_{\alpha\beta}
= - \frac{3 i M \bar\rho^4}{8} \!\! \fourint{l} A(l) F(l) 
\left( \Gamma_\alpha \gamma_\beta \tau U^{\gamma_5} + U^{\gamma_5} \tau \gamma_\beta \Gamma_\alpha \right). \!
\label{contraction_projected}
\end{align}
The final integral evaluates to
\begin{align}
\bar\rho^4 \fourint{l} \, A(l) F(l) &= \frac{2}{3}.
\end{align}
This remarkable relation follows from the Dirac equation for the zero mode wave function
in the instanton field in momentum representation \cite{Balla:1997hf}.
Eq.~(\ref{contraction_projected}) reduces to
\begin{align}
\Gamma_{\alpha\beta}
&= - \frac{i M}{4}
\left( \Gamma_\alpha \gamma_\beta \, \tau U^{\gamma_5} + U^{\gamma_5} \tau \, \gamma_\beta \Gamma_\alpha \right) ,
\end{align}
which is independent of the instanton size $\bar\rho$.
This vertex represents the effect of the gauge potential in the operator Eq.~(\ref{operator_euclidean})
in the instanton vacuum.
Altogether, including also the derivative term in Eq.~(\ref{operator_euclidean}), we find that the
QCD operator Eq.~(\ref{operator_euclidean}) can be represented by the effective two-fermion operator
\begin{align}
(O_\Gamma)_{\alpha\beta}(x) &=
-i \psi^\dagger (x) \left\{ \gamma_{\alpha} \overleftrightarrow{\partial}_\beta \tau
\right.
\nonumber
\\
& \left. - \frac{i M}{4}
\left[ \Gamma_\alpha \gamma_\beta \, \tau U^{\gamma_5}(x) + U^{\gamma_5}(x) \tau \, \gamma_\beta \Gamma_\alpha \right]
\right\} \psi (x).
\label{onebody_final}
\end{align}
This two-fermion operator gives the same result as the four-fermion operator Eq.~(\ref{effective_operator})
at the parametrically leading level when inserted in hadronic correlation functions with external momenta
$p_{1, 2} \sim M \ll \bar\rho^{-1}$. Here we have restored the position dependence of the background field
$U^{\gamma_5}(x)$ on the scale $M \ll \bar\rho^{-1}$. Equation~(\ref{onebody_final}) is the main result of
present work and has numerous implications for hadron structure.

We observe that the instanton induces a ``potential'' term in the effective operator, which accompanies
the ``kinetic'' term of the quark field derivatives. The potential term is of the order of the dynamical
quark mass term in the effective action Eq.~(\ref{S_eff}) and therefore has a large effect on hadron structure.
It is chirally odd and therefore involves the chiral background field (this is required by chiral symmetry).
It has a specific spin/flavor dependence conditioned by the symmetries of the instanton field.
Altogether, this highlights how instantons convert color dynamics to an effective spin-flavor dynamics.

The instanton effect depends on the spin or twist of the operator. In the twist-2 projection of
Eq.~(\ref{onebody_final})
(symmetric traceless tensor) the potential term is absent. This happens because for
$\Gamma_\alpha = \gamma_\alpha$ the instanton-induced vertex in Eq.~(\ref{onebody_final}) involves
$\gamma_\alpha \gamma_\beta$ and $\gamma_\beta\gamma_\alpha$, which cannot produce a symmetric traceless tensor,
\begin{align}
\gamma_\alpha \gamma_\beta + \gamma_\beta \gamma_\alpha - \textrm{trace} = 0;
\end{align}
the same happens for $\Gamma_\alpha = \gamma_\alpha \gamma_5$. This generalizes to twist-2
spin-$n$ operators \cite{Balla:1997hf}. The twist-2 quark distributions in the instanton vacuum
therefore can be calculated with the ``kinetic'' operators in leading order of the packing fraction.
The consistency of this approximation is shown by the fact that the partonic sum rules (charge, momentum)
are satisfied at this level \cite{Diakonov:1996sr}.
In the twist-3 projection of Eq.~(\ref{onebody_final}) (antisymmetric tensor) the potential term is present.
In this case the products $\gamma_\alpha \gamma_\beta$ and $\gamma_\beta\gamma_\alpha$ in the instanton-induced
vertex project on $\sigma_{\alpha\beta}$ and give rise to non-zero structures. Regarding the twist-4
projection (trace) of Eq.~(\ref{onebody_final}), see Footnote~\ref{footnote:vev}.
\section{Twist-3 effective operators}
Using the master formula Eq.~(\ref{onebody_final}) we can now obtain the effective two-fermion operators
for the twist-3 QCD operators Eqs.~(\ref{operator_flavor_natural}) and (\ref{operator_flavor_unnatural}).
It is convenient to present the effective operators in Minkowskian form, so that they can be compared
directly to the original QCD operators. For the natural-parity operator
Eq.~(\ref{operator_flavor_natural}) we obtain 
\begin{align}
O^{\alpha\beta} (x)
&= \bar\psi (x) \left\{ \frac{1}{2} \gamma^{[\alpha} i \overleftrightarrow{\partial}^{\beta]} \tau
+ \frac{iM}{4} \sigma^{\alpha\beta} [\tau , U^{\gamma_5}(x) ] \right\} \psi (x),
\label{twist3_natural}
\end{align}
where we have used $\sigma^{\alpha\beta} U^{\gamma_5} = U^{\gamma_5} \sigma^{\alpha\beta}$. The potential
term is proportional to the flavor commutator $[\tau , U^{\gamma_5}]
\equiv \tau U^{\gamma_5} - U^{\gamma_5}\tau$. It is therefore zero in the flavor-singlet
case $\tau = 1$, which means that there is no instanton effect (or chiral field effect) on the total quark spin
density. It is non-zero in the flavor-nonsinglet case $\tau \neq 1$, which means that there is an instanton
effect on the individual flavor distributions of the quark spin. It is instructive to exhibit this effect in terms
of the conventional pion field. Expanding the chiral field Eq.~(\ref{U_gamma_5}) in the pion field,
\begin{align}
U^{\gamma_5}(x) &= 1 + i \gamma_5 \pi^b(x) \tau^b /F_\pi + ...,
\end{align}
the operator Eq.~(\ref{twist3_natural}) with $\tau = \tau^a$ becomes
\begin{align}
\bar\psi (x) \left\{ \frac{1}{2} \gamma^{[\alpha} i \overleftrightarrow{\partial}^{\beta]} \tau^a
- \frac{i M}{2 F_\pi} \sigma^{\alpha\beta} \gamma_5 \, \epsilon^{abc} \pi^b (x) \tau^c \right\} \psi (x).
\end{align}
One sees that the potential term couples the pion field to the isovector pseudotensor
current of the quark field.

For the unnatural-parity operator Eq.~(\ref{operator_flavor_unnatural}) we obtain
\begin{align}
& O_5^{\alpha\beta} (x)
\nonumber
\\
&= \bar\psi (x) \left( \frac{1}{2} \gamma^{[\alpha} \gamma_5 \, i \overleftrightarrow{\partial}^{\beta]} \tau
- \frac{iM}{4} \sigma^{\alpha\beta} \gamma_5 \{ \tau, U^{\gamma_5}(x) \} \right) \psi (x).
\label{twist3_unnatural}
\end{align}
Here the potential term is proportional to the flavor anticommutator
$\{\tau , U^{\gamma_5}\} \equiv \tau U^{\gamma_5} + U^{\gamma_5}\tau$
and affects both flavor singlet and nonsinglet quark spin-orbit correlations.
\section{Effective equations of motion}
In the effective theory obtained from instantons, the quark fields are subject to equations of motion
in the chiral background field, governed by the effective action Eq.~(\ref{S_eff}).
These effective equations of motion imply certain relations between the operators in the effective theory.
When applied to the twist-3 effective operators, the effective equations of motion reproduce the QCD
operator relations Eqs.~(\ref{eom_natural}) and (\ref{eom_unnatural}). In Minkowskian form, the
equations of motion for quark fields in the effective theory Eq.~(\ref{S_eff}) are
\begin{align}
[ i \overrightarrow{\slashed{\partial}} - M U^{\gamma_5}(x)] \psi (x) = 0,
\hspace{1em}
\bar\psi (x) [ -i \overleftarrow{\slashed{\partial}} - M U^{\gamma_5}(x)]  = 0.
\label{eom_effective}
\end{align}
Here we consider energies/momenta $p \sim M \ll \bar\rho^{-1}$, which is sufficient for the typical fields
in hadron structure. Using Eqs.~(\ref{eom_effective}), one can easily show that the Minkowskian
twist-3 effective operator Eq.~(\ref{twist3_natural}) obeys the relation
\begin{align}
&\bar\psi (x) \left\{ \frac{1}{2} \gamma^{[\alpha} i \overleftrightarrow{\partial}^{\beta]} \tau
+ \frac{iM}{4} \sigma^{\alpha\beta} [\tau , U^{\gamma_5}(x) ] \right\} \psi (x)
\nonumber \\
&= -\frac{1}{4}
\epsilon^{\alpha\beta\gamma\delta} \partial_\gamma \left[\psi (x) \gamma_\delta \gamma_5 \tau \psi (x) \right],
\label{eom_effective_natural}
\end{align}
where the operator on the R.H.S is the axial current in the effective theory. The relation holds for any
$\tau$ (singlet and nonsinglet). The instanton-induced potential term in the effective operator
is essential for bringing about Eq.~(\ref{eom_effective_natural}); without it there would be
a discrepancy of the order of the dynamical quark mass. The same is obtained for the unnatural-parity
twist-3 effective operator Eq.~(\ref{twist3_unnatural}).
This remarkable result comes about because (i) in leading order of the packing fraction the gauge potential
in the QCD operator is represented by a single instanton; (ii) the quark fields in the instanton background
satisfy the QCD equations of motion (the zero mode is a solution to the Dirac equation in the instanton field).
It attests to the consistency of the approximation scheme based on the packing fraction expansion
(diluteness) and the $1/N_c$ expansion (saddle point approximation).

An important practical consequence of the effective operator relation Eq.~(\ref{eom_effective_natural})
is that the hadron structure results do not depend on which ``version'' of the operator is used.
This problem afflicts dynamical models without a systematic derivation of the effective dynamics
and the effective operators.
\section{Quark spin-orbit correlations}
Our findings have consequences for the phenomenology of quark spin-orbit correlations in the nucleon
described by the matrix elements of the twist-3 operator
Eq.~(\ref{operator_flavor_unnatural}) \cite{Lorce:2014mxa}. A detailed study of the nucleon
matrix elements in the mean-field picture based on the instanton-induced
effective theory and the $1/N_c$ expansion will be presented elsewhere. Here we limit ourselves to
two limiting cases of this picture, the ``quark model'' and ``chiral soliton'' limits
(their relation to the general mean-field picture is discussed e.g. in Ref.~\cite{Praszalowicz:1995vi}).
They illustrate the impact of the instanton-induced potential in the effective operator and provide
a baseline for further detailed studies.

In the quark model limit, we consider the matrix element of the effective operator Eq.~(\ref{twist3_unnatural})
in a weakly bound system of massive quarks ($U^{\gamma_5} \equiv 1$).
The essential point can be seen by taking the operator Eq.~(\ref{twist3_unnatural})
with $\tau = 1$ and $U^{\gamma_5} \equiv 1$,
\begin{align}
& O_5^{\alpha\beta} (x)
= \bar\psi (x) \left( \frac{1}{2} \gamma^{[\alpha} \gamma_5 \, i \overleftrightarrow{\partial}^{\beta]}
- \frac{iM}{2} \sigma^{\alpha\beta} \gamma_5 \right) \psi (x),
\label{twist3_quarkmodel_unnatural}
\end{align}
and computing the expectation value in a free quark state
with Minkowskian 4-momentum $p$ and $p^2 = M^2$, Dirac spinor wave function
$u$ satisfying $(\slashed{p} - M) u = 0$, and polarization vector
$s^\alpha \equiv \bar u \gamma^\alpha \gamma_5 u$. The matrix element of the kinetic term is
\begin{align}
\frac{1}{2} \langle p | \gamma^{[\alpha} \gamma_5 i \overleftrightarrow{\partial}^{\beta ]} |p \rangle
= s^{[\alpha} p^{\beta ]} .
\end{align}
The matrix element of the potential term is obtained as
\begin{align}
-\frac{i M}{2}
\langle p | \sigma^{\alpha\beta} \gamma_5 |p \rangle
=  -s^{[\alpha} p^{\beta ]}.
\end{align}
The matrix element of the complete effective operator is then
\begin{align}
\langle p | O_5^{\alpha\beta} |p \rangle = 0;
\end{align}
this could also be inferred from the fact that the complete effective operator is a total derivative.
One sees that the instanton-induced potential term cancels the trivial result
in the free state arising from the kinetic term and makes the the spin-orbit correlation a pure interaction effect.

In the ``chiral soliton'' limit, the quark fields are integrated out, and we consider the nucleon as
a soliton of the chiral Lagrangian Eq.~(\ref{chiral_action}). The effective operator corresponding to this
situation is obtained from Eq.~(\ref{twist3_unnatural}) by averaging over the quark fields in the
background of the chiral fields,
\begin{align}
& O_5^{\alpha\beta} [U] \equiv \langle O_5^{\alpha\beta} [\bar\psi, \psi] \rangle_U .
\label{chiral_unnatural}
\end{align}
The average can be computed using the Green function of the quark field in the background of the chiral
field (loop diagram). The dependence on the chiral field can be made explicit by gradient expansion
in chiral field, similar to the effective action Eq.~(\ref{chiral_action});
the technique is described in Refs.~\cite{Diakonov:1987ty,Diakonov:1996sr}.
For example, the operator for the isovector vector current in the chiral theory is obtained as (here $N_f = 2$)
\begin{align}
& \langle \bar\psi \gamma^\alpha \tau^a \psi \rangle_U
= \frac{F_\pi^2}{2i}\textrm{tr}\left[
(U^\dagger \partial^\alpha U +  U \partial^\alpha U^\dagger ) \tau^a \right]
+ \mathcal{O} (\partial U^3);
\end{align}
this result can also be obtained by applying the symmetry transformation to the
leading-order term of the chiral Lagrangian Eq.~(\ref{chiral_action}) \cite{Zahed:1986qz}.
For the isovector twist-3 operator Eq.~(\ref{chiral_unnatural}) ($\tau = \tau^a$),
the gradient expansion gives
\begin{align}
& \langle O_5^{\alpha\beta} [\tau = \tau^a] \rangle_U
\nonumber \\
&= -\frac{F_\pi^2}{8 i} \epsilon^{\alpha\beta\gamma\delta} \partial_\gamma
\, \textrm{tr}\left[
(U^\dagger \partial_\delta U +  U \partial_\delta U^\dagger) \tau^a \right]
+ \mathcal{O} (\partial U^4) ,
\end{align}
which is the total derivative of the isovector current of the chiral field and agrees
with the equation-of-motion relation Eq.~(\ref{eom_unnatural}).
In this case we observe that the leading-order result in $\partial U$ arises from the kinetic term in the
operator Eq.~(\ref{twist3_unnatural}); the contributions from the potential term
come in only at higher orders. For the isoscalar twist-3 operator Eq.~(\ref{chiral_unnatural}) ($\tau = 1$),
the gradient expansion has to be carried out including terms of order $\partial U^4$,
as the isoscalar vector current of the chiral field is of order $\partial U^3$ \cite{Zahed:1986qz};
the expressions will be presented elsewhere.
\section{Summary and extensions}
Instantons induce potential terms in the twist-3 quark operators measuring the quark spin density and spin-orbit
correlations in hadrons. The potential terms are proportional to the dynamical quark mass emerging from chiral
symmetry breaking, are independent of the instanton size, and have a distinctive spin-flavor structure.
They ensure that the effective operators obey the same equation-of-motion relations as the original QCD operators.
The potential terms qualitatively change the spin-orbit correlations in hadrons compared to estimates
with the kinetic terms only.

Our results provide the basis for a systematic analysis of twist-3 GPDs in the instanton vacuum.
The present study has focused on a special case of twist-3 operators that are related to well-known
twist-2 operators through the equation-of-motion, which has allowed us to test the consistency
of the approximations based on the instanton packing fraction and the $1/N_c$ expansion.
Future studies can address cases where such simple relations are not available and one has to
rely on the approximations to organize the dynamics.

The methods developed here can be extended to several other cases of interest: (i)
higher moments of the chirally-even twist-3 GPDs; (ii) chirally-odd twist-3 GPDs;
(iii) matrix elements of twist-4 operators.

The present study employs the minimal description of the QCD vacuum in terms of instanton-type fluctuations,
which induce fermionic zero modes and cause chiral symmetry breaking. Recent work has explored a more extended
description including other types of topological fluctuations (so-called molecules and zig-zag paths),
which do not possess zero modes but contribute to observables such as Wilson lines;
see Refs.~\cite{Shuryak:2021fsu,Shuryak:2021hng} and other articles in the series.
Extending the effective operator approach to this more general framework would be an interesting problem
for further study.
\section*{Acknowledgments}
This material is based upon work supported by the U.S.~Department of Energy, 
Office of Science, Office of Nuclear Physics under contract DE-AC05-06OR23177.
The research reported here takes place in the context of the Topical Collaboration
``3D quark-gluon structure of hadrons: mass, spin, tomography'' (Quark-Gluon Tomography Collaboration)
supported by the U.S.~Department of Energy, Office of Science, Office of Nuclear Physics.

\bibliography{instanton}
\end{document}